\documentclass[aps,prl,amsmath,amssymb,preprintnumbers,superscriptaddress,footinbib,twocolumn,showpacs,floatfix]{revtex4}
\usepackage{epsfig,graphics,color}
\usepackage{graphicx}
\usepackage{dcolumn}
\usepackage{bm}
\usepackage{amsmath}

\newcommand \ra  {\rightarrow}

\newcommand \h {\theta}

\newcommand \A {\alpha}

\newcommand \lc {\langle}
\newcommand \rc {\rangle}
\newcommand \prt {\partial}

\newcommand \sg {\sigma}

\newcommand \bvec{\left( \begin{array}{c} }
\newcommand \evec{\end{array} \right)}

\newcommand \bea{\begin{eqnarray} }
\newcommand \eea{\end{eqnarray} }
\newcommand \nn {\nonumber}
\newcommand {\be} {\begin{equation}}
\newcommand {\ee} {\end{equation}}

\newcommand {\mbx} {\mbox{}}

\newcommand {\ata} {& \times &}

\begin{document}

\title{Suppression of the high $p_T$ charged hadron $R_{AA}$ at the LHC} 

\author{A.~Majumder}
\affiliation{Department of Physics, The Ohio State University, Columbus, OH 43210, USA}
\affiliation{Department of Physics and Astronomy, Wayne State University, Detroit, MI 48201, USA}

\author{C.~Shen}
\affiliation{Department of Physics, The Ohio State University, Columbus, OH 43210, USA}

\date{ \today}

\begin{abstract}
We present a parameter free postdiction of the high-$p_T$ 
charged-hadron nuclear modification factor ($R_{AA}$) in two centralities, measured by the CMS collaboration in 
$Pb$-$Pb$ collisions at the Large Hadron Collider (LHC). 
The evolution of the bulk medium is modeled using viscous fluid dynamics, with parameters adjusted to 
describe the soft hadron yields and elliptic flow.
Assuming the dominance of radiative energy loss, we compute the medium modification of the $R_{AA}$ 
using a perturbative QCD based formalism, the higher twist scheme. The transverse momentum diffusion coefficient $\hat{q}$ 
is assumed to scale with the entropy density and normalized by fitting the $R_{AA}$ in the most central $Au$-$Au$
collisions at the Relativistic Heavy-Ion Collider (RHIC). This set up is validated in non-central $Au$-$Au$ collisions at 
RHIC and then extrapolated to $Pb$-$Pb$ collisions at the LHC, keeping the relation between $\hat{q}$ and entropy density  
unchanged. We obtain a satisfactory description of the CMS $R_{AA}$ over the $p_{T}$ range from $10$-$100$ GeV. 
\end{abstract}


\maketitle



Jet quenching, or the modification of hard jets in a dense 
extended medium, is one of the most studied discoveries at 
the Relativistic Heavy-Ion Collider (RHIC)~\cite{RHIC_Whitepapers}. 
Numerous experiments have established the suppression of 
high transverse momentum (high $p_T$) hadrons~\cite{highpt} 
at RHIC energies.
There now exist volumes of theoretical calculations based on 
both perturbative QCD (pQCD)~\cite{Majumder:2010qh} and the AdS/CFT conjecture~\cite{Gubser:2009sn} aiming to 
describe jet modification at RHIC. 
Collisions of heavy-ions at the LHC, as measured by all three 
detectors (ATLAS, ALICE, CMS), have demonstrated  significant evidence for the modification of hard jets~\cite{Aad:2010bu,Aamodt:2010jd,CMS:2012aa}.

Before any such measurement may be applied to the detailed study of the structure of the dense medium,  
theoretical calculations must be able to describe the basic jet quenching observables.
A standard jet quenching measure, as established by the RHIC experiments, is the 
nuclear modification factor $R_{AA}$, defined as, 
\bea
R_{AA} = \frac{ \frac{ d^{2} N_{AA} ( b_{min}, b_{max} )  }{dp_{T}^{2} dy }   } 
{  N_{bin} ( b_{min}, b_{max} )  \frac{d^{2} N_{pp} }{ dp_{T}^{2} dy } }. \label{RAA}
\eea
In the equation above, $d^{2} N_{AA}(b_{min}, b_{max})$ represents the yield of hadrons in 
narrow bins of $p_{T}$, rapidity ($y$) and centrality (designated by a range of impact parameters 
$ b_{min}$ to $b_{max}$) of the heavy-ion collision. 
In the denominator, $N_{bin}$ represents the number of binary nucleon-nucleon collisions in the
same centrality bin and $d^{2} N_{pp}$  represents the yield of hadrons in $p$-$p$ collisions, in the same 
bin of $p_{T}$ and $y$. So far both ALICE and CMS have reported the $R_{AA}$ of charged hadrons at the 
LHC~\cite{Aamodt:2010jd,CMS:2012aa}.

In this Letter, we present a parameter free comparison with the 0-5\% and 10-30\% CMS data~\cite{CMS:2012aa}. 
The calculation consists of two factorized parts: 
A fluid dynamical simulation whose initial conditions and transport coefficients were
tuned to describe the hadron spectrum and elliptic flow at $p_{T}\!\! < \!\!2$ GeV in $Au$-$Au$ collisions at RHIC and 
successfully extrapolated to $Pb$-$Pb$ collisions at the LHC~\cite{Shen:2011}, and a pQCD 
based radiative energy loss calculation which computes the medium modified spectrum of 
high $p_{T}$ hadrons. Unlike soft hadrons, these hard hadrons will be assumed to stem from the fragmentation of hard jets, 
modified due to passage through the medium.

By ``parameter free'' we mean that the transverse momentum diffusion coefficient $\hat{q}$, which controls the 
amount of radiative energy loss encountered by a hard jet, will be related to intrinsic quantities in the 
fluid dynamics simulation which are entirely controlled by the soft hadron obsrvables. 
In this case, it will be scaled with the temperature dependent entropy density $s$:
\bea
\hat{q}(\vec{r},t) = \hat{q}_{0} s(T(\vec{r},t)) \Big/\left[ s(T_{0})\sqrt{1 - v_\perp^2(\vec{r},t)} \right] , \label{qhat}
\eea
where $q_{0}$ refers to the value of the transport coefficient at the highest RHIC temperature of $T_{0} = 344$ MeV, 
reached at the center of the 0-5\% central $Au$-$Au$ collisions at $0.6$ fm/c, and $v_\perp$ refers to the local flow velocity transverse to the jet. Once related to temperature in this way, $\hat{q}$ 
is completely controlled by the fluid dynamics simulation at any point in space-time, in any collision at any energy or centrality. 
Given Eq.~\eqref{qhat}, 
the space-time dependent entropy density in the 
simulation for LHC energies will predict the value of $\hat{q}$ at any space-time point.

The remaining Letter is organized as follows: after a brief review of the essential features of the higher twist method to describe
 parton energy loss, we discuss specific ingredients of the current calculation. This will be followed by 
 a comparison to $R_{AA}$ in central and semi-central collisions at RHIC (i.e. $Au$-$Au$ at $\sqrt{s} = 200$ GeV/nucleon),
 as well as the in and out of plane $R_{AA}$ in semi-central RHIC  
 events; $\hat{q}_{0}$ will be dialed to obtain a best fit to this data. Following this the $R_{AA}$ will be computed 
 for 0-5\% and 10-30\% centrality events at the LHC (i.e. $Pb$-$Pb$ at $\sqrt{s} = 2.76$ TeV). We will conclude with a discussion of 
 possible reasons why the theory underpredicts the $R_{AA}$ at $p_{T}\leq 8 $ GeV at the LHC.

It will be assumed that processes such as the production of high $p_{T}$ hadrons which 
engender a hard scale $(Q \gg \Lambda_{QCD})$ through much of their space-time evolution may 
be computed using pQCD. 
Furthermore, the hard scale allows for the use of factorization, effectively separating soft subprocesses from the part where the scales involved are hard~\cite{Collins:1985ue}. 
In such a factorized formalism, the invariant cross section to produce a high $p_{T}$ hadron in a heavy-ion collision 
may be expressed as 
\bea
\mbx\!\!\!\frac{d \sigma^{AA}(p_T,y)_{b_{min}}^{b_{max}} }{dy d^2 p_T} 
\!\!\! &=& K \int_{\mbx_{b_{min}}}^{\mbx^{b_{max}}}\!\!\!\!\!\!\!\!\!d^2 b 
\int d^2 r t_A(\vec{r}) t_B(\vec{r} - \vec{b})  \nn \\
\!\!\! &\times& \!\!\!\! \int  d x_a d x_b  G^A_a(x_a,Q^2)  G^B_b(x_b,Q^2)  \nn \\
\ata \!\!\!\! \frac{  d \hat{\sg}_{ab \ra cd} }{ d \hat{t}}  
\frac{ \tilde{D}_c^h(z,Q^2,\zeta_{L}(\vec{r},\theta_{j}), \hat{p}_{c} )}{\pi z}.  \label{AA_sigma}
\eea
In the equation above, $y,\vec{p}_{T}$ represents the rapidity and transverse momentum of the detected 
parton, $K$ is a multiplicative factor to account for higher order corrections, 
$\vec{r}$ is the location of the jet production vertex, $t_{A/B} (\vec{r} \pm \vec{b}/2 ) $ is the nuclear 
thickness function [$t_{A} (\vec{r} + \vec{b}/2) = \int dz \rho(\vec{r} + \vec{b}/2, z) $ with $\rho(\vec{r},z)$ the nucleon 
density inside a nucleus], $x_{a}(x_{b})$ represents the momentum 
fractions of the incoming partons, $Q^{2} \sim p_{T}^{2}$ is the hard scale of the process, 
$G_{a/b}^{A/B}(x_{a}, Q^{2})$ is the nuclear parton distribution function, 
$d \hat{\sg}_{ab \ra cd}/ d \hat{t}$ is the short distance cross section for incoming partons $a,b$ to 
scatter and produce partons $c,d$ with a squared momentum transfer $\hat{t} = (\hat{p}_{c} -  \hat{p}_{a} )^{2}$, 
$\tilde{D} $ represents the medium modified fragmentation with hadronic momentum fraction 
$z=p_{T}/{p_{c}}_{T}$, and $\zeta_{L}(r,\theta_{j})$ is the distance travelled by a jet produced at 
$\vec{r}$ and propagating at an angle $\theta_{j}$ with respect to the reaction plane. Both $\vec{b}$ and $\vec{r}$ are two dimensional 
vectors transverse to the beam direction. 
All calculations will be 
carried out at mid-rapidity  ($y = 0$).

In this effort  
we will ignore 
shadowing corrections, both at LHC and at RHIC. This will give the 
calculated RHIC $R_{AA}$ a slightly rising slope with $p_{T}$,
compared to previous calculations. The medium modified fragmentation 
function is calculated by solving an in-medium DGLAP evolution equation~\cite{Majumder:2009zu} valid when 
$Q^{2} \gg \hat{q} \tau_{f}$ (where, $\tau_{f} \sim p/Q^{2}$ is the lifetime of a parton with energy $p$ and virtuality 
$Q^{2}$, undergoing a split),  
\bea
\mbx &&\!\!\!\!\!\! \!\!\!\!\!\!\frac{ \prt  \! \tilde{D}_c^h(z,Q^2, \hat{p}_{c} ) }{ \prt\log{Q^{2} } } 
\!\!\!\!\!\mbx^{ \mbx^{\mbox{ \scalebox{1} { $ |_{\vec{r}}^{\vec{r} + \hat{n}\zeta_{L} } $ } } } }
\!\!\!\!\!\!\!\!\!\!\!=\! \frac{\A_s}{2\pi} \int\limits_z^1 \frac{dy}{y} \tilde{P}(y) 
\left[ \tilde{D}_c^h\left(\frac{z}{y},Q^2\!\!, \hat{p}_{c} \right)_{\vec{r}}^{\vec{r} + \hat{n}\zeta_{L} } \right. \nn \\
&+& \left. \int\limits_{\vec{r}}^{ \vec{r} + \hat{n} \zeta_{L} }\!\!\!\!\! d\zeta  K_{\hat{p}_{c},Q^2}^{\vec{r},\theta_{j} } ( y,\zeta)  
 {D_q^h}\left. \left(\frac{z}{y},Q^2\!\!,\hat{p}_{c} y \right) \right|_{\vec{r} + \hat{n} \zeta}^{\vec{r} + \hat{n} \zeta_L}  \right] , 
 \label{in_medium_evol_eqn}
\eea
where, $\tilde{P}(y)$ is the Altarelli-Parisi splitting function for a parton $c$ to split into two other partons which 
carry $y$ and $1-y$ fractions of the momentum of $c$.
The in-medium single emission kernel $K_{\hat{p}_{c},Q^2}^{\vec{r},\theta_{j} } $~\cite{HT} is given as 
\bea
K_{\hat{p}_{c},Q^2}^{\vec{r},\theta_{j} }  = \frac{\hat{q}}{Q^{2}}(\vec{r} + \hat{n} \zeta)
\left[ 2 - 2 \cos\left\{ \frac{Q^2 (\zeta )}{ 2 \hat{p}_{c} y (1- y)} \right\}  \right] , \label{kernel}
\eea
involving the space-time dependent transport coefficient $\hat{q}$ introduced in Eq.~\eqref{qhat}.

As stated in Eq.~\eqref{qhat}, the jet transport parameter depends on the local entropy density $s$. 
The entropy density is obtained from a $2+1$D viscous fluid dynamical simulation where the input parameters, 
such as the magnitude of the components of the initial energy momentum tensor as well as the viscosity and
the final freezeout criteria are tuned to obtain the best fit with the spectra and elliptic flow of hadrons with 
$p_{T} \leq 2$ GeV. As this Letter will deal solely with the spectra and azimuthal anisotropy at high 
$p_{T}$ ($p_{T} \geq 6$ GeV), we will only provide the most salient features of these simulations in the current paper and 
direct the reader to Refs.~\cite{Song:2010mg,Shen:2010uy}
  for further details. 
  
In the fluid dynamical simulations 
used, the thermalization time is $\tau_{0} \!=\! 0.6\!$~fm/c for both RHIC and LHC collisions. 
Initial conditions are based on the KLN parametrization of the collision of two nuclei with saturated gluon distributions~\cite{Song:2010mg}.
The jets are assumed to be produced according to a binary collision profile at the moment of 
collision, i.e. at $t_{0}=0$. For times between $t_{0}$ and $\tau_{0}$, the 
conditions at $\tau_{0}$ are extrapolated back to $t_{0} = 0$ fm/c unchanged, i.e. we assume that the soft medium 
remains unchanged from $0$ to $0.6$ fm/c.

Like prior calculations of jet modification in a fluid dynamical set up, the hadronic 
phase is also simulated as a viscous fluid and not in terms of a hadronic cascade. While such hybrid calculations 
are currently available, we have picked the simplest simulation for this attempt to calculate the $R_{AA}$ at the 
LHC. As we use the entropy density to scale the local value of $\hat{q}$, there is no extra rescaling factor for the 
hadronic phase as in prior attempts where  $\hat{q}$ was scaled with $T^{3}$~\cite{Bass:2008rv,Majumder:2007ae}.
The produced jets are assumed to decouple from the medium when the local temperature reaches $160\!$~MeV.
Given the above conditions, a space-time profile for $\hat{q}$ is obtained for jets traveling in all directions, 
starting at $t=0$ at any location in the medium, and vanishing
when $T(\vec{r},t )  = 160\!$~MeV. This profile is completely 
specified with the value of $\hat{q}_{0} \! = \! 2.2\!$~GeV$^{2}$/fm at $T\! =\!344\!$~MeV which is the highest temperature reached at RHIC, 
at $\vec{r} = 0$ in the 0-5\% most central collisions at $t = \tau_{0}$. This $\hat{q}_{0}$ is obtained by requiring that the 
computed $R_{AA}$ in 0-5\% central collisions at $p_{T} = 10\!$~GeV  is approximately $0.2$.

Using this $\hat{q}$ profile, one can compute 
the medium modified fragmentation function $\tilde{D}$ [Eq.\eqref{in_medium_evol_eqn}] for jets which originate at $\vec{r}$ 
and travel in a given direction $\theta_{j} (\hat{n})$, decoupling from the medium after traveling a distance 
$\zeta_{L} (\vec{r}, \theta_{j})$ [when $T(\vec{r}+\hat{n} \zeta_L, \zeta_L) = 160\!$~MeV],
with a given initial energy $\hat{p}_{c}$, and a final hadron momentum specified in terms of $p_{T}$ and $y$. 
The medium modified fragmentation function is obtained by solving the differential equation in Eq.~\eqref{in_medium_evol_eqn} 
starting with an input vacuum fragmentation function at $Q^2 = Q_{min}^2$, and evolving up to $Q^2 = p_T^2$. 
The physical picture is that of a hard jet that starts with a high virtuality $Q^2 = p_T^2$ and drops down in 
virtuality as it splits into less virtual partons while propagating through the medium. Eventually at a $Q^2= Q_{min}^2$ it 
exits the medium. Thus, there is a strong correlation between energy, virtuality and distance travelled by the jet.

For this attempt to predict the $R_{AA}$ at the LHC, we will make two sets of approximations which will include the 
correlation between energy, virtuality and position in terms of their averages, thus providing a systematic error in 
extrapolating the calculations at a given $p_{T}$, centrality and energy of collision to other systems.
First, we integrate the product of the energy loss 
kernel $K$ in Eq.~\eqref{kernel} and the nuclear overlap function [$T_{AA} (\vec{b}) = \int d^{2} r t_{A}(\vec{r} - \vec{b}/2) t_{B} (\vec{r} + \vec{b}/2)$] over all jet origins and directions of exit, for all values of the ratio $Q^{2}/(2 \hat{p}_{c})$.
Note that the energy loss kernel correlates location with energy $\hat{p}_{c}$ and virtuality $Q^{2}$ through this ratio.
We then carry out the evolution in virtuality $Q^{2}$ for jets with all allowed energies 
using the position and jet direction integrated kernel. For each $p_{T}$ of the detected hadron,  
$Q_{min}^{2} = \lc \hat{p}_{c} \rc/\lc \zeta_{L} \rc$ where $\lc \hat{p}_{c} \rc$ is the mean energy of the 
parent parton in vacuum and $\lc \zeta_{L} \rc$ is the mean distance travelled by the jets that escape the medium 
with a virtuality $Q^{2} \geq Q_{min}^{2}$.
For one set of curves (dashed lines in all plots), 
the mean $\hat{q}$ of the medium is calculated by averaging over all possible jet paths in the fluid medium. Then a single 
emission formalism (in a homogeneous static medium with the mean $\hat{q}$) is used to calculate the length $\zeta_{max}$ 
over which a parton with energy $\lc \hat{p}_{c} \rc$ will lose all its energy.
The mean path length  $\lc \zeta_{L} \rc$ is then calculated by averaging over paths in the fluid medium with $\zeta_{max}$ as the maximum allowed length.
Since a single emission is somewhat inefficient in removing the energy from the parton, this yields a larger $\lc \zeta_{L} \rc$ 
leading to smaller $Q_{min}^{2}$ and thus an over estimate of the quenching. 
For hadron $p_{T} \sim 10$ GeV in 0-5\% collisions at RHIC, $\zeta_{max}  = 5$~fm. For the other set of curves, 
(solid lines in all plots) $\lc \zeta_{L} \rc$ is calculated with the restriction that all lengths larger than 5 fm are excluded. 
 Since jets with larger energy should penetrate deeper in the medium, 
this provides an underestimate of the quenching.

\begin{figure}[!htbp]
\resizebox{3.0in}{2.5in}{\includegraphics{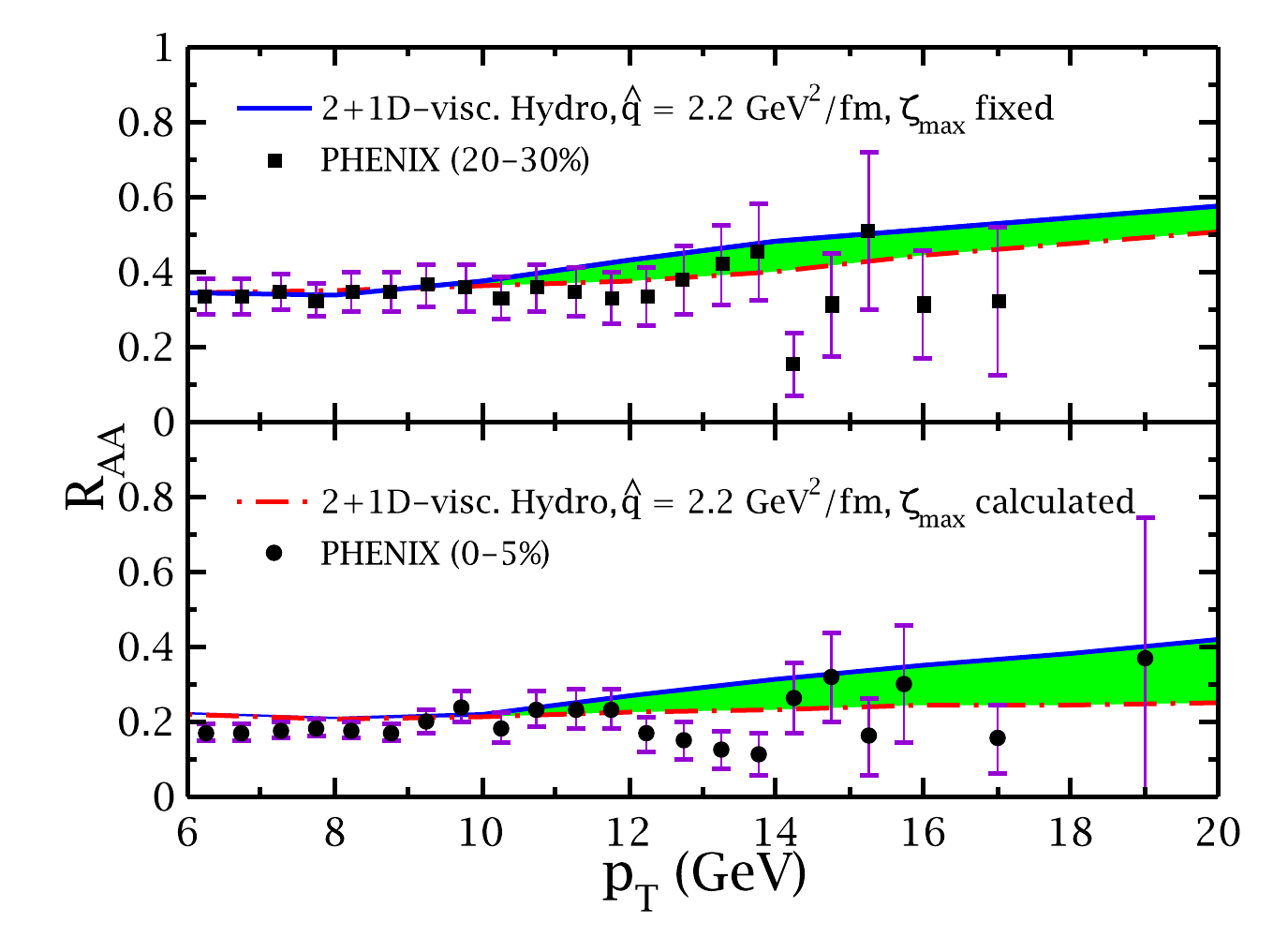}} 
\vspace{-0.25cm}
    \caption{(Color online) The $R_{AA}$  versus $p_{T}$  for two centralities.  
    The green bands represent the calculations presented in this Letter with a $\hat{q}_{0} = 2.2$~GeV$^{2}$/fm. 
    No shadowing is included and the minimum virtuality ($Q_{min}^{2}$) is set to 
    ${\rm max}\{ \lc \hat{p}_{c} \rc/\lc \zeta_{L} \rc, 1 {\rm GeV}^{2} \}$.  See text for details.} 
    \label{fig2}
\label{raa-centrality}
\end{figure}

\begin{figure}[htbp]
\resizebox{2.5in}{2.5in}{\includegraphics{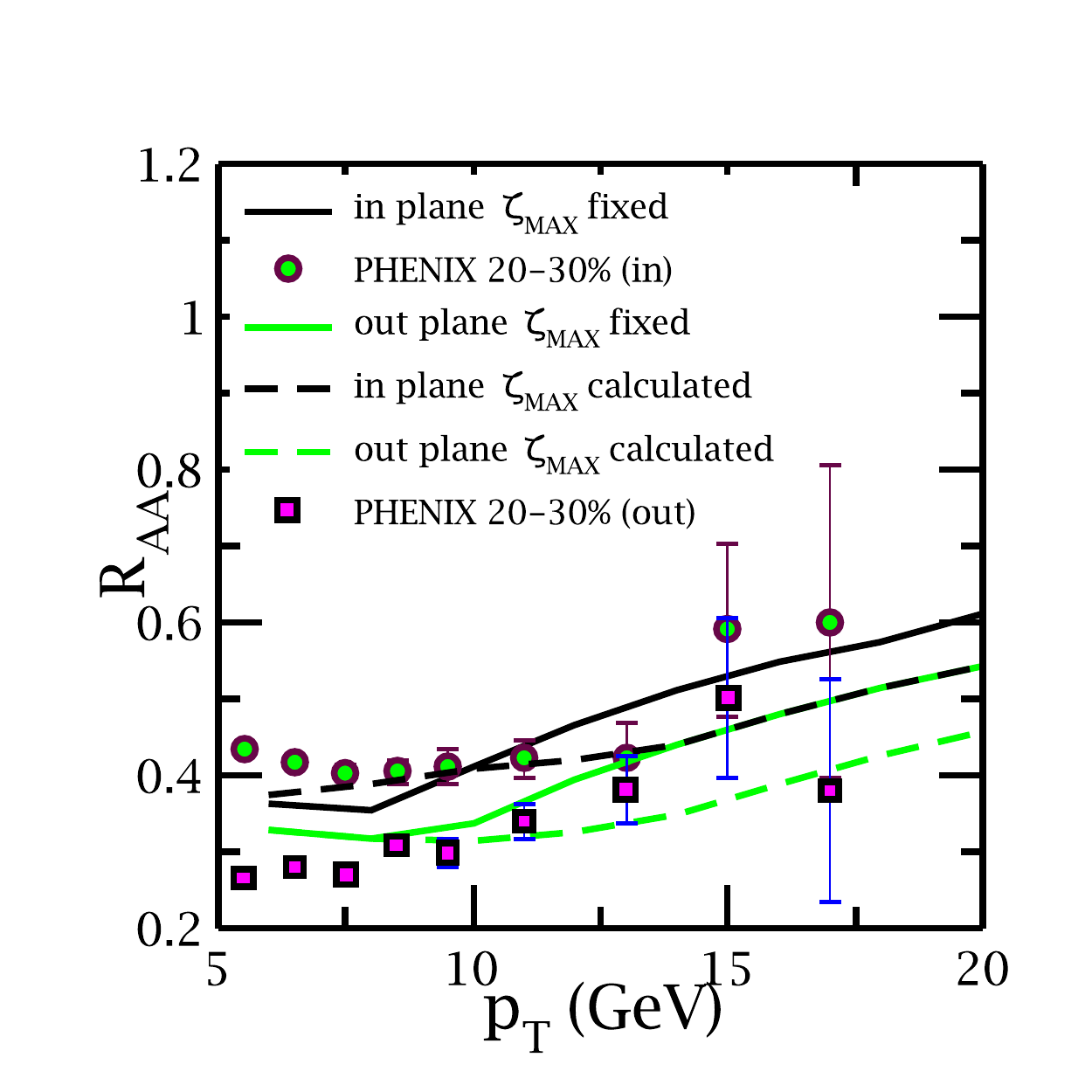}} 
\vspace{-0.25cm}
    \caption{(Color Online) A parameter free calculation of the $R_{AA}$ in plane and out of plane in the 
    20-30\% centrality collisions and comparison with PHENIX data for the same. } 
    \label{in-out-plane}
\end{figure}

With this set up, we can now calculate the invariant cross section to produce a high $p_{T}$ hadron in a heavy-ion 
collision using Eq.~\eqref{AA_sigma} and obtain the $R_{AA}$ using Eq.~\eqref{RAA}. 
The results for RHIC are shown in Fig.~\ref{raa-centrality}, 
compared with  the 
$R_{AA}$ measured by PHENIX in two centrality bins (0-5\% and 20-30\%). As mentioned above, 
the one dimensionfull parameter $\hat{q}_{0}$ is dialed so that the calculated $R_{AA}$ matches the experimental value 
in the 0-5\% collisions at $p_{T} = 10$ GeV [i.e. $R_{AA} (0-5\%,p_{T} = 10 \mbox{GeV}) \simeq 0.2$].
The slope of the curve with $p_{T}$ and the shift with centrality are predictions. The $\hat{q}_{0}$ required is 
$2.2$ GeV$^{2}$/fm. 
In these calculations, $Q^{2}_{min} = \lc \hat{p}_{c} \rc/\lc \zeta_{L} \rc$ until this value drops below 
1 GeV$^{2}$ wherein we hold it fixed at 1 GeV$^{2}$ (this occurs at a $p_{T} \lesssim 8$~GeV). 
Our choice of $\hat{q}_{0}$ is validated by comparing with the in-plane and out-of-plane $R_{AA}$
in the 20-30\% centrality collisions as measured by PHENIX (plotted in Fig.~\ref{in-out-plane}). 
The solid black curve represents the $R_{AA}$ for jets with $\theta_{j}$ chosen such that $0 < \theta_{j} < \pi/12$, 
for a case of a fixed $\lc \zeta_{max}\rc$. 
Note, $\h_{j} = 0$ is the direction of $\vec{b}$.
The solid green line is the $R_{AA}$ with $5\pi/12<\theta_{j}<\pi/2$, for a case of a fixed $\lc \zeta_{max}\rc$. The dashed  
black and green lines are the same calculations but with calculated $\lc \zeta_{max}\rc$.
We obtain a good description of the data for $p_{T}\geq 8$ GeV. For $p_{T} < 8$~GeV, the $Q^{2}_{min}\geq 1$~GeV$^{2}$ restriction begins to 
affect the calculation.

\begin{figure}[htbp]
\resizebox{3in}{2.5in}{\includegraphics{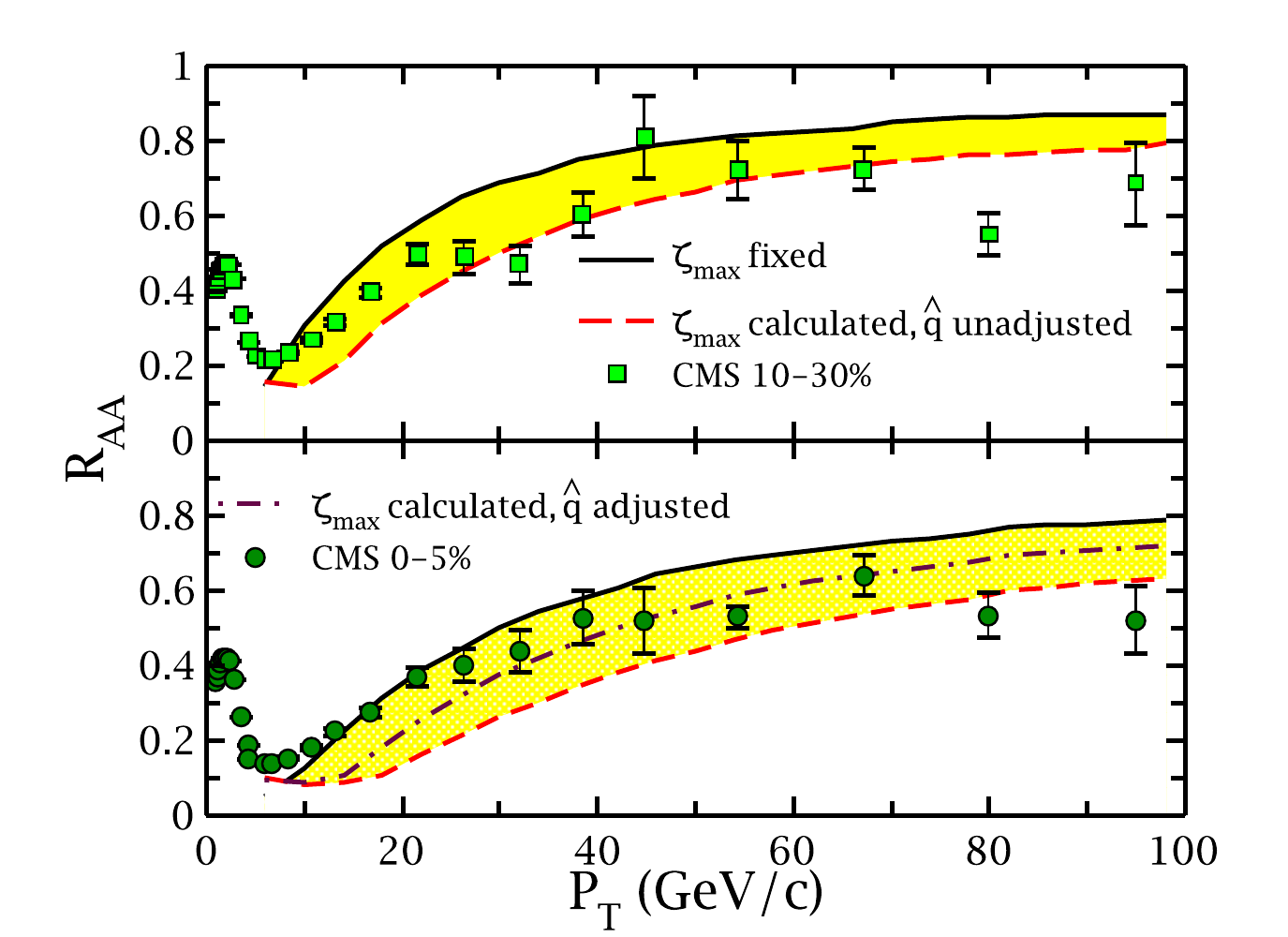}} 
\vspace{-0.25cm}
    \caption{(Color Online) A parameter free calculation of the $R_{AA}$ in the 
    0-5\% most central and 20-30\% central collisions, and comparison with CMS data for the same. } 
    \label{fig3}
\end{figure}

With the value of $\hat{q}_{0}$ extracted from RHIC collisions and $\hat{q}$ scaled with the entropy density of the 
medium we now calculate the $R_{AA}$ in 
$Pb$-$Pb$ collisions at $\sqrt{s}\! =\! 2.76$~ATeV, 
at the LHC. 
In order to fit the 
increased charged hadron multiplicities, 
one requires the fluid dynamical simulation to possess almost twice as much entropy density 
at thermalization as at RHIC ($T_{max}\! = 436$ MeV). Also larger velocity gradients drive the system to larger sizes than at RHIC 
prior to freezeout~\cite{Shen:2011}. As a result, the mean length traversed by jets escaping the medium ($\lc \zeta_{L} \rc$) is larger.
To obtain the $R_{AA}$ we 
perform an almost identical calculation as that for RHIC energies, the sole difference being the range of available jet 
energies. As shown in Fig.~\ref{fig3}, we obtain a good description of the measured $R_{AA}$ (for both 0-5\% and 10-30\% centralities) for $p_{T} \!\geq\! 9$ GeV. 
The solid and dashed lines indicate the cases of fixed and calculated $\lc \zeta_{max}\rc$ as before. 
This indicates that, in spite of the 
produced medium being both larger and denser, the basic jet quenching observables at high $p_{T}$ at the LHC can be described 
in an identical pQCD based formalism as used at RHIC. 

No doubt, future efforts will improve on various approximations made in this 
calculation, in particular the treatment of the correlation between energy, virtuality and distance travelled by a parton. However it is not 
clear if these corrections are the only source of the discrepancy between the calculation and the data at $p_{T} \sim 6$~GeV.
It is entirely possible that non-perturbative phenomena such as recombination extend up to 8-9 GeV at the LHC, due to the larger 
flow at these energies. The extraordinarily high densities may also enhance the magnitude of non-linear effects in jet quenching which 
have so far been ignored. In order to gauge this uncertainty, we also plot the $R_{AA}$ with a calculated $\lc \zeta_{max}\rc$ and 
adjust the $\hat{q}_{0}$ to obtain a better fit with the 0-5\% LHC data. This seems to require a 30\% lower value of $\hat{q}_{0}$.
Another source of correction is the drag and diffusion experienced by the partons~\cite{Majumder:2008zg}, not included in this calculation

In conclusion, we have performed a pQCD based calculation of the suppression of the high $p_{T}$ hadron spectra due to 
radiative energy loss of hard partons at both 
RHIC and LHC. 
The local transport coefficient $\hat{q}$ which controls the magnitude of radiative energy loss is scaled with the 
entropy density.
The entropy density profile of the produced matter is controlled by a 2+1D viscous fluid dynamical simulation which 
has been tuned to describe the soft spectrum and elliptic flow. The overall normalization of $\hat{q}$ is provided by 
setting its value at a given $T$ or $s$, in this case $\hat{q}_{0} \!=\! 2.2\,{\rm GeV}^{2}$/fm at the maximum RHIC temperature of 
$T\!=\! 344\!$~MeV. With this value being set, we predict the $p_{T}$ dependence of the $R_{AA}$ in 0-5\% and 20-30\% centrality 
 events at RHIC as well as the $R_{AA}$ versus reaction plane, and the $R_{AA}$ measured in 0-5\% 
and 20-30\% centrality events at $\sqrt{s} = 2.76\!$~ATeV at the LHC.

The authors thank U.~Heinz, Y.~Kovchegov, M.~A.~Lisa and B.~M\"{u}ller for discussions, and the WSU grid for computer time.
This work was supported in part by the U.S. DOE under grant nos. DE-SC0004286 and (within the framework of the JET 
collaboration) DE-SC0004104, and in part by the NSF under grant no PHY-1207918.

\end{document}